# Optical properties of two-dimensional tin nanosheets epitaxially grown on graphene

Eleonora Bonaventura[1], Christian Martella[1], Salvatore Macis[2], Daya S. Dhungana[1], Simonas Krotkus [4], Michael Heuken[4], Stefano Lupi[2-3], Alessandro Molle[1*], Carlo Grazianetti[1*]

[1]CNR-IMM Unit of Agrate Brianza, via C. Olivetti 2, Agrate Brianza, IT
[2]Department of Physics, Sapienza University, Piazzale Aldo Moro 5, Roma, IT
[3]CNR-IOM, Q2 Building, Area Science Park, Basovizza-Trieste, IT
[4]AIXTRON SE, Dornkaulstraße, Herzogenrath, GE

*Corresponding Authors: alessandro.molle@cnr.it (A.M.), carlo.grazianetti@cnr.it (C.G.)

**Abstract**

Heterostacks formed by combining two-dimensional materials show novel properties which are of great interest for new applications in electronics, photonics and even twistronics, the new emerging field born after the outstanding discoveries on twisted graphene. Here, we report the direct growth of tin nanosheets at the two-dimensional limit via molecular beam epitaxy on chemical vapor deposited graphene on $Al_2O_3$(0001). The mutual interaction between the tin nanosheets and graphene is evidenced by structural and chemical investigations. On the one hand, Raman spectroscopy indicates that graphene undergoes compressive strain after the tin growth, while no charge transfer is observed. On the other hand, chemical analysis shows that tin nanosheets interaction with sapphire is mediated by graphene avoiding the tin oxidation occurring in the direct growth on this substrate. Remarkably, optical measurements show that the absorption of tin nanosheets show a graphene-like behavior with a strong absorption in the ultraviolet photon energy range, therein resulting in a different optical response compared to tin nanosheets on bare sapphire. The optical properties of tin nanosheets therefore represent an open and flexible playground for the absorption of light in a broad range of the electromagnetic spectrum and technologically relevant applications for photon harvesting and sensors.

1. Introduction

The rise of two-dimensional vertical heterostructures is currently reshaping the frontiers of new materials research. The advantages of assembling different layers include not only an increase in possible functionalities, but also a higher stability of the individual heterostructure components and Xenes, the post-graphene synthetic lattices, are not excluded from this promising perspective [1,2]. The first reported examples are of environmentally stable Xenes, such as phosphorene, mechanically exfoliated and stacked with other two-dimensional materials [3,4]. Recently, the use of stanene as a template for the epitaxial growth of silicene (and vice versa) has proven to be a viable route for the realization of Xene-based heterostructures [5]. Moreover, the addition of a layer of stanene has been shown to weaken the interaction between the silicene and its native Ag substrate, making the properties of such decoupled silicene more accessible [6]. Epitaxial growth methods allow for the fabrication of heterostructures on a large scale, overcoming the limitations of mechanically assembled van der Waals flakes, whose lateral dimensions do not exceed tens of microns [7]. Despite these advances, metallic substrates severely limit the Xenes device integration and this makes insulators particularly attractive for electronic and photonic applications [8–10].

The use of c-plane sapphire or $Al_2O_3(0001)$ as a template is a suitable choice for the experimental realization of Xenes [11,12]. However, although there is no evidence of interaction with the substrate for epitaxial silicon nanosheets, different results were obtained for tin. The optical properties of tin nanosheets grown directly on $Al_2O_3(0001)$ by molecular beam epitaxy (MBE) showed stanene-like optical features when compared to calculated absorption spectra of freestanding stanene [12–14]. These stanene-like features deduced from optical spectroscopy were observed even if the chemical analysis carried out by X-ray photoelectron spectroscopy (XPS) demonstrated that tin undergoes a partial oxidation [12]. The formation of tin oxide has been explained by the unavoidable interaction between tin and the adsorbed hydroxyl group (-OH) on the surface of the $Al_2O_3(0001)$ substrate [15,16]. Engineering the substrate for the Xenes growth therefore has a twofold purpose. First, given the intrinsic metastable nature of Xenes, it is crucial to control the interaction between the material of interest and the supporting substrate. Second, from a general perspective, stacking different two-dimensional materials offers a great potential for tailoring the physical and chemical properties of each constituent or the whole coupled system [17]. Graphene on $Al_2O_3$ is a technologically relevant configuration that has attracted considerable interest in recent years [18–20]. The reduction of metal contamination, unintentional doping and mechanical stress are some of the benefits of using sapphire as a target substrate. In addition, the atomically planar graphene surface can act as a template to stabilize the growth of various two-dimensional materials [21,22]. Few works show the direct synthesis of Xenes on graphene to obtain vertical hetero-stacks, for example silicene [23–25]. Instead, the possibility of growing stanene on graphene has been explored in different theoretical works [26,27], while its experimental realization has recently been reported on Cu(111), achieving a uniform thickness, continuous morphology, and higher stability [28].

Here, we report on the use of graphene $Al_2O_3(0001)$ as a template for the epitaxial growth of tin nanosheets at the two-dimensional limit. The structural properties of the heterostructure were investigated by Raman spectroscopy while optical spectroscopy was then used to measure the optical response of the tin nanosheets in the photon energy range from 0.3 to 6.5 eV (from NIR to UV). The outcome is an open playground where the optical properties of the whole system can be easily modified, allowing versatile tuning of light absorption, especially when these findings are compared with the case without graphene in between tin nanosheets and

Al$_2$O$_3$(0001).

## 2. Materials and methods

High-quality monolayer graphene grown on single side polished *Al$_2$O$_3$*(0001) has been used as substrate for tin nanosheets deposition. The 2-inch original wafer, provided by Aixtron, has been cut into ∼1×1 $cm^2$ pieces by a cleaving tool (LatticeGear). After the cut, graphene layers quality was evaluated via Raman spectroscopy and AFM analysis. Tin deposition on graphene was performed by MBE on degassed samples using a k-cell evaporator for which the flux rate was previously calibrated though a quartz micro-balance. Two tin thicknesses were considered, 0.5 and 1 nm, and were grown at room temperature. Both types of samples were then capped by an amorphous aluminum oxide capping layer (∼5 nm) in order to prevent their degradation in ambient conditions [29]. The overall stacking is thus (from top to bottom) a-Al$_2$O$_3$/Sn/Gr/Al$_2$O$_3$(0001). Chemical properties of the samples were investigated *in situ* by means of XPS, monitoring the status of graphene and tin films at all stages of the growth. Non-monochromatized Mg Kα X-ray source was used to investigate the C 1s and Sn 3d$_{5/2}$ core levels. Raman spectroscopy was used to assess quality, doping and strain of graphene before and after the tin deposition. Raman maps over areas of 7 × 7 µm$^2$ were acquired in backscattering configuration using a Renishaw InVia spectrometer equipped with 514 nm laser and 50× (N. A. 0.75) objective. The incident laser power was kept below 3 mW to avoid heating effects and sample damages. The optical characterization was performed in the NIR-UV spectral range. The transmittance was measured at room temperature using a Cary 5000 spectrophotometer, covering the energy range between 0.3 and 6.5 eV. Oxygen plasma etching was used to remove graphene from the sapphire surface to provide a good reference for optical characterization. The optical conductivity of thin tin films was extracted though RefFIT by implementing a multilayer model [41]. Specifically, the transmittance of the samples, including the substrate, was fitted by using a double (graphene-*Al$_2$O$_3$*) or a three (Sn-graphene-*Al$_2$O$_3$*) layers Kramers-Kronig consistent Drude-Lorentz model.

## 3. Results and discussion

Single-layer graphene was grown on Al$_2$O$_3$(0001) by chemical vapor deposition (CVD), while tin deposition on graphene on Al$_2$O$_3$(0001) (hereafter Gr-Al$_2$O$_3$) was then performed by MBE at room temperature, in an ultra-high vacuum (UHV) environment (see Methods). Two different tin nanosheet thicknesses, 0.5 and 1 nm, were considered in the following. *In situ* XPS was carried out to check the chemical status of the tin films after the growth. The Sn 3d$_{5/2}$ core level, for the 0.5 nm-thick sample, is reported in **Figure 1a.** The spectrum shows the core level related to the elemental tin placed at binding energy (BE) of 484.7 eV with a full-width half maximum (FWHM) of 1.15 eV. When the tin thickness is increased to 1nm, the Sn 3d$_{5/2}$ core level in **Figure 1b** is still characterized by a single component at the same BE. The absence of any oxide-related component indicates that graphene is effective in avoiding the chemical interaction between the tin nanosheets and the *Al$_2$O$_3$*(0001) substrate. Furthermore, the lack of additional components in the Sn 3d$_{5/2}$ core level spectrum suggests that the chemical interaction between tin and graphene is low, as also confirmed by the C 1s spectra reported in **Figure A.1** of Supplementary Information. **Figure A.1** shows that the C 1s core level is made of two components, i.e. one related to $sp^2$ carbon (graphene) and the other to C-H bonding [19]. After tin deposition, the graphene-related component of the C 1s XPS spectrum undergoes a negligible BE shift (+0.11 and -0.11 eV, for 0.5 and 1 nm, respectively), thus making evidence of limited chemical interaction between the tin nanosheets and the underlying graphene.

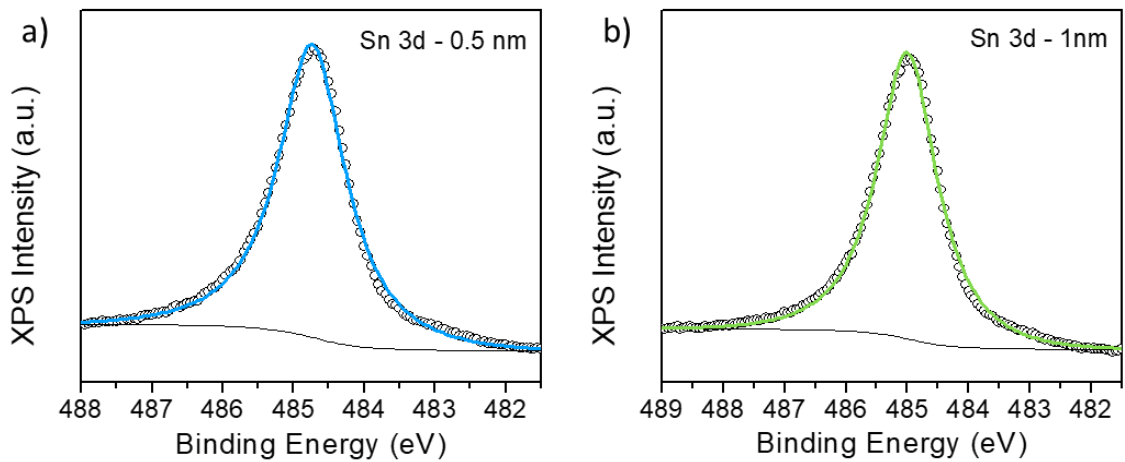

**Figure 1**: XPS spectra of the Sn $3d_{5/2}$ core level of 0.5 (blue) and 1 (green) nm tin grown at RT on Gr-$Al_2O_3$ showing a single component related to the elemental states of Sn. Open circles represent the raw data.

The samples were then capped with a 5 nm-thick amorphous $Al_2O_3$ layer grown *in situ* by MBE to prevent oxidation in ambient conditions and to allow for Raman and optical spectroscopy investigations (see Materials and methods) [29].

Tin exhibits active Raman modes for wavenumbers below 300 cm$^{-1}$ in both its bulk and two-dimensional forms [30–32]. Raman characterization of the tin nanosheets, in the 100 to 400 cm$^{-1}$ range, revealed no peaks associated with tin as already observed for the case without graphene because of the low Raman scattering cross-section of tin in this nanoscale regime [12]. However, unlike direct growth on $Al_2O_3$(0001), the graphene buffer layer allows us to infer how the presence of tin nanosheets affects the underlying graphene by modifying its well-known Raman modes [33]. Indeed, by evaluating the effects of tin nanosheet growth on the Raman response of graphene, it was possible to gain information about the properties of the whole heterostructure. In fact, as a first approximation, for a single layer of graphene, any shift or line widening of the G and 2D Raman modes can be related to strain and/or doping effects [34]. The Raman spectrum of graphene on $Al_2O_3$(0001) obtained from a single-point acquisition is shown in **Figure A.2a** where the fingerprints of the electron-phonon interaction in a single layer graphene are disclosed. The statistical analysis, carried out on Raman maps acquired over several areas of 7 × 7 µm$^2$, results in the G band located at 1588 ± 1 cm$^{-1}$ with a FWHM of 18 ± 3 cm$^{-1}$ and the 2D band at 2690 ± 2 cm$^{-1}$ with the FWHM of 39 ± 4 cm$^{-1}$. These values are in agreement with those measured for graphene directly grown on $Al_2O_3$(0001) and reported in works using a similar statistical approach [18,35,36]. Moreover, the values of the 2D/G intensity ratio 2.5±0.2 indicate a low charge carrier concentration [37]. According to the calibration curve reported by Das et al. [37] we estimate a carrier concentration of ∼ $1.5 \times 10^{12}$ $cm^{-2}$. **Table A.1** in Supplementary Information reports the values obtained from the statistical analysis performed on the three areas of the wafer highlighted (green squares) in **Figure A.2c**. The surface morphology of graphene on $Al_2O_3$ (0001) in **Figure A.2b**, obtained by atomic force microscopy (AFM), shows a high density of wrinkles on the slabs surface with a typical height of 1-5 nm. Such structures are consequence of the mismatch between the thermal expansion coefficient of graphene and sapphire as reported in literature [18,19]. With the graphene modes kept as reference, we investigated the evolution of the Raman spectrum of graphene with increasing thickness of the tin nanosheets (**Figure 2a**).

With respect to the Raman modes of the pristine graphene, after the tin deposition we observed the broadening of the 2D mode and the blue-shift of both G and 2D modes. Specifically, the FWHM of the 2D peak in the spectra reported in **Figure 2a** increases with the thickness of the deposited tin, going from 39 to 42 $cm^{-1}$ for 0.5 and 1 nm-thick nanosheets, respectively. The 2D/G peaks intensity ratio, despite a slight increase to 2.7 ± 0.2, is compatible with the value obtained for bare graphene and with a negligible (or low) charge carrier transfer (i.e. doping, see also discussion below) from both 0.5 and 1 nm deposited tin [37]. Moreover, the D peak is still visible without any significant variation suggesting that the homogeneity of graphene has not been affected by the growth.

As mentioned above, Raman spectroscopy is a very helpful method for extracting information about the doping and strain to which graphene is subjected. An inverse approach was therefore used to shed light on the interaction between tin and graphene in the stacked configuration. We exploited the variation of the Raman response of graphene, induced by the deposition of tin films with different thicknesses, with the aim of gaining information on the resulting heterostructure. The correlation plot of the Raman shifts of the G and 2D peaks, shown in **Figure 2b**, allows us to separate the contribution of doping and strain [34]. The two dashed orange lines specify the directions along which the strain-induced and the hole/doping-induced shifts are expected. Moreover, the point where the two lines intersect, located at ($\omega^0_G$, $\omega^0_{2D}$) = (1582, 2677), identifies graphene not affected by strain or excess charges [34]. The correlation map shows that graphene on the $Al_2O_3$(0001) substrate is strained prior to tin deposition. This effect can be due to the interaction between graphene and sapphire, which results in the formation of the wrinkles. No substantial differences are observed between the uncovered graphene and after the deposition of the 0.5 nm-thick tin nanosheet. Conversely, when tin nanosheets thickness increases, the dispersion shows that graphene undergoes a compressive strain while the effect on charge doping is negligible. Assuming a uniaxial strain-sensitivity of the G mode $\Delta\omega_G/\Delta\varepsilon$ = −23.5 $cm^{-1}$/% the corresponding compressive strain due to the 1 nm-thick nanosheet deposition is about 0.5% [34,38]. Interestingly, the type of stress to which graphene is subjected can provide information about the possible intercalation of tin atoms. It has recently been reported that under certain temperature and impurity conditions, silicene grown on graphene can exhibit intercalation [39]. In this case, the Raman spectrum of graphene shows a splitting of the G peak and the 2D peak downshift, both aspects being indicative of the tensile strain experienced by graphene [40]. The different Raman response and the compressive strain to which graphene is subjected after the tin nanosheets deposition suggests ruling out intercalation effects. Moreover, it is important to point out that the effects such as desorption and intercalation typically occur during high-temperature growth or post-growth annealing, whereas all the samples scrutinized here were grown at room temperature.

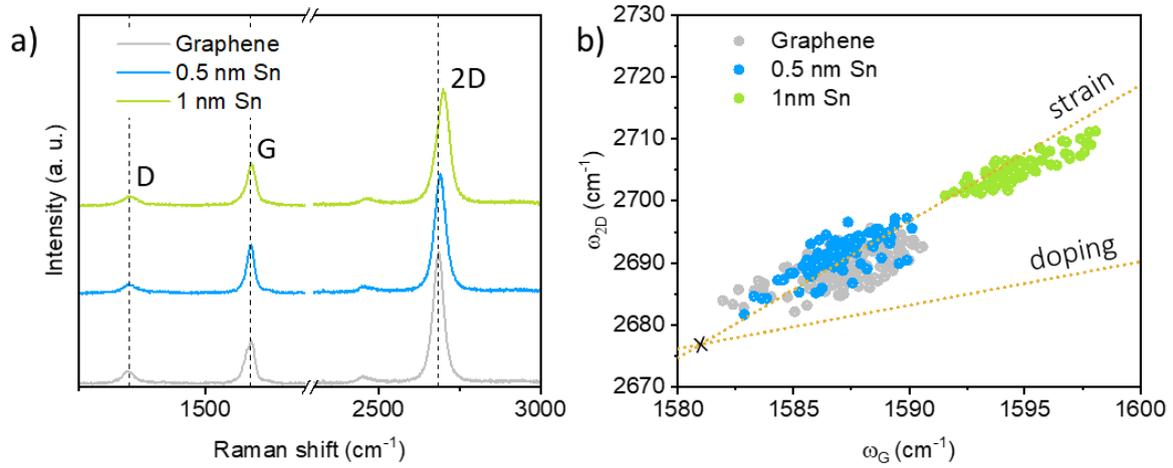

**Figure 2**: a) Raman spectra of graphene before (gray) and after the growth of tin nanosheets 0.5 nm (blue) and 1 nm (green) thick.  b) Correlation plot between the frequencies of the G and the 2D modes of graphene ($\omega_G$, $\omega_{2D}$) measured on pristine graphene and after the growth of tin nanosheets, 0.5 nm (blue) and 1 nm (green) thick. The "strain" orange line describes neutral-charge graphene under compressive or tensile strain. The "doping" orange line indicates p-type doped graphene.

The optical response of the heterostructures formed by the tin nanosheets and graphene on $Al_2O_3(0001)$ has been investigated by NIR-UV spectroscopy. **Figure 3a** shows the absolute transmittance for the two tin nanosheets samples, together with those for Gr-$Al_2O_3$ and the bare $Al_2O_3(0001)$, in the photon energy range between 0.3 and 6.5 eV. As expected, the observed decrease of transmittance in the spectra of **Figure 3a** is consistent with an increase in absorption in the thicker tin nanosheets. The optical conductivity of graphene, 0.5 nm- and 1 nm-thick tin nanosheets was obtained by fitting (red-dashed line in **Figure 3a**) the transmittance spectra by means of the RefFIT software (see Materials and methods) [41]. The extracted real part of the optical conductivity $\sigma_1(\omega)$ was used to calculate the normalized optical conductance $G_1/G_0 = \sigma_1(\omega)d/G_0$ reported in **Figure 3b**, where $d$ is the thickness of the considered layer and $G_0 = e^2/4\hbar$ is the universal conductance [42,43]. The optical conductance can be used for a fair comparison between samples of different thicknesses. $G_1(\omega)$ for graphene shows the typical behavior described by the superposition of a nearly flat background in the whole IR range (0.3-1.5 eV) due to the intraband transitions, and an asymmetric absorption peak at ~4.6 eV [44]. This feature is related, in the independent-particle description, to the interband transitions from the bonding to the antibonding $\pi$ states near the saddle-point singularity at the M point of Brillouin zone [45].

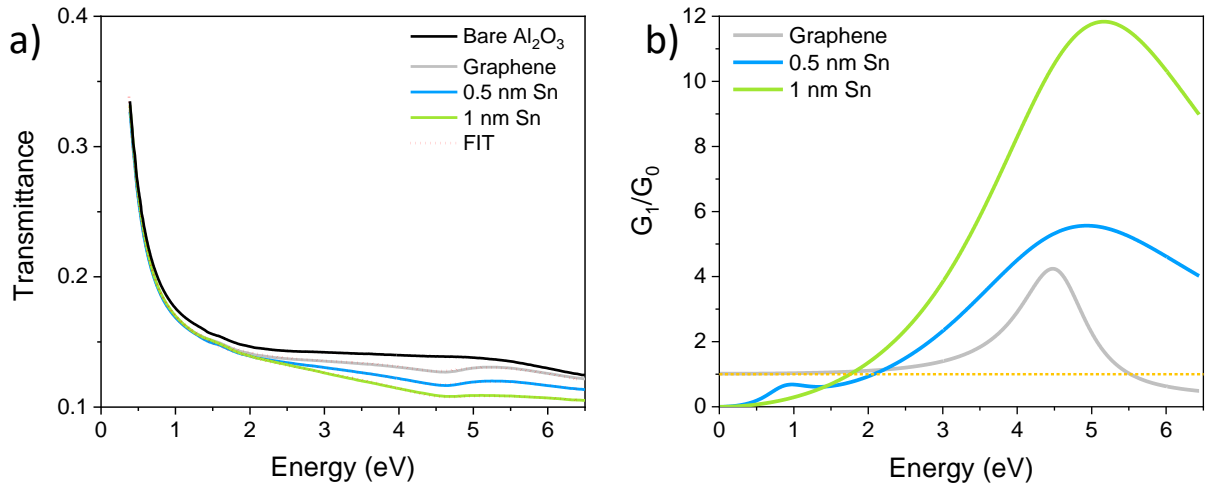

**Figure 3**: a) Transmittance of supported graphene (gray), 0.5 nm (blue) and 1 nm (green) thick tin nanosheets. The bare *Al₂O₃*(0001) transmittance (black line) is reported as reference. b) Normalized optical conductance of graphene (gray), 0.5 nm (blue) and 1 nm (green) thick tin nanosheets, extracted from multi-layer models.

At the lowest thickness of tin, $G_1(\omega)$ shows a small absorption peak around 0.9 eV, without the flat background, and rises around 1.5 eV up to the UV range (blue curve in **Figure 3b**). The maximum value occurs at ∼5 eV shifted and broadened with respect to graphene absorption. When the tin nanosheet thickness increases to 1 nm, the absorption peak in the IR region disappears and the rise becomes steeper with a small shift of the $G_1(\omega)$ maximum to further higher energies. Surprisingly, both tin nanosheets exhibit an enhanced optical absorption compared to graphene, starting at ∼2 eV, as expected for stanene in the stanene-graphene heterostructure [27]. It should be emphasized that the real part of the optical conductivity, and therefore the conductance, shown in **Figure 3b**, is related to the individual tin layers. In the applied multilayer model, the contribution of graphene and substrate can be considered separately, allowing the optical response of the tin nanosheets to be extracted from the overall optical response of the whole system. For the sake of completeness, the latter one is shown in **Figure A.3**. As the thickness of the tin layer increases, the optical contribution of graphene to the shape of the overall system response decreases, making the optical response of the combined tin nanosheets-graphene system strongly tin-like.

As already shown for the direct deposition on Al₂O₃(0001) [12], the optical behavior of the tin nanosheets is unusual even with the graphene buffer layer. This aspect is supported by the comparisons between the normalized optical absorbance A(ω) of different tin-based structures, shown in **Figure 4**. The absorbance was calculated from the imaginary part of the refractive index ñ(ω)=n(ω)+iκ(ω) as A(ω)=2κ(ω)ωd, where α(ω)=2κ(ω)ω is the absorption coefficient of the material. The absorption spectra of the 0.5 and 1 nm-thick tin nanosheets reported in **Figure 4a** deviate from those of the two allotropic forms of tin, the α- and β-Sn [46,47]. At low frequencies, however, their A(ω) approaches zero similar to α-Sn. In this context, it is worth noting that the epitaxy on lattice-matched substrates has been reported as a possible strategy for stabilizing the α-Sn at room temperature [30,48]. Moreover, it is important to stress that the poor correspondence between the optical response of the tin nanosheets and the α-Sn might be attributed to the small thickness of the nanosheets compared to the reference (128 nm) [46]. **Figure 4b** shows the comparison between A(ω) of the tin nanosheets and the calculated absorbance of free-standing stanene reported in Refs. [13] and [14]. The spectrum calculated by John et al. [14] shows low- and high-energy features that match with the experimental data for the broad peaks at ~1 and ~5 eV. Finally, **Figure 4c** shows the absorption spectrum of the 0.5 nm-thick tin nanosheet grown on Gr- Al$_2$O$_3$ and that of a tin nanosheet of the same thickness directly grown on the Al$_2$O$_3$ substrate [12]. The spectra in **Figure 4c** are qualified by a shift of the absorption maximum towards higher photon energy when tin is grown on a buffer layer of graphene. The Lorentzian oscillator used to model this absorption moves from 2.8 to 4.9 eV in the presence of graphene. Similarly, the second Lorentzian contribution, which is used to model the absorption at lower energies, shifts down from 1.3 eV for direct tin growth to 0.9 eV with graphene. We can thus conclude that the interlayer coupling plays a crucial role in modifying the optical properties of a hybrid system. The optical response of the tin nanosheets supported by Al$_2$O$_3$(0001) or Gr-Al$_2$O$_3$(0001) can be tuned over a wide range of photon energies covering the VIS and the UV spectrum. Such a tunability of the optical conductivity can be achieved by differently assembling the elements of the heterostructure stacking.

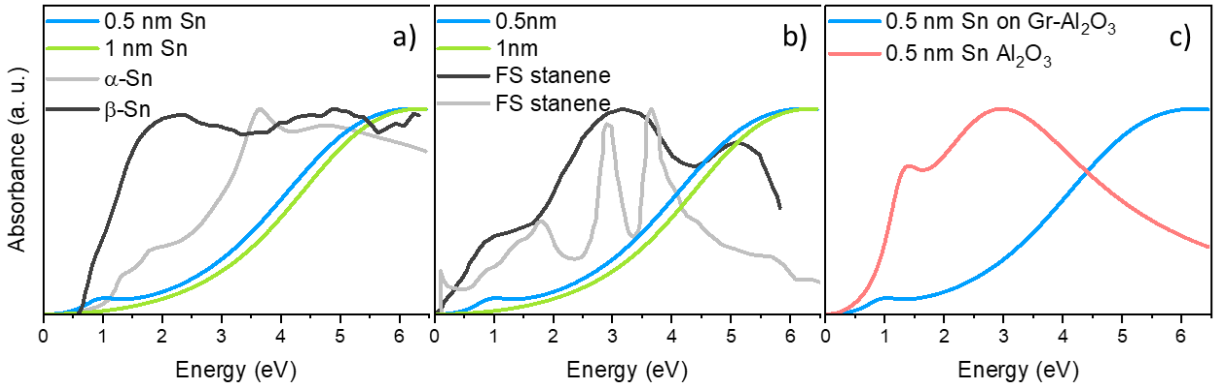

**Figure 4:** a) Absorption spectra of 0.5 nm- (blue) and 1 nm-thick (green) tin nanosheets and those of α-Sn (gray) and β-Sn (black) (data from Refs. [46,47]. b) Absorbance of 0.5 nm- (blue) and 1 nm-thick (green) tin nanosheets compared with the absorbance of freestanding stanene reported in Ref. [13] (gray) and Ref. [14] (black). c) Absorbance of 0.5 nm-thick tin nanosheet grown on Gr-Al$_2$O$_3$ (blue) and 0.5 nm-thick tin nanosheet grown directly on Al$_2$O$_3$ (red) reported in Ref. [12].

## 4. Conclusions

The successful growth of tin nanosheets at two-dimensional limit has been reported on the Gr-Al$_2$O$_3$(0001) template. At variance with the direct growth on Al$_2$O$_3$(0001), inserting a graphene buffer layer in between turns out to minimize the chemical interaction with the substrate and then the oxidized component. The structural characterization by Raman spectroscopy demonstrated that graphene after tin deposition is affected by a tin thickness-dependent compressive strain but not by charge transfer effects, i.e. doping. The strain can be related to the interaction with tin overlayer which, however, shows no signs of intercalation or chemical bonding between tin and carbon. Interestingly, the role of graphene is not limited to the prevention of tin oxidation, but also enables the optical engineering of the tin nanosheets. Optical measurements in the NIR-UV spectral range identify an increased absorption starting from 2 eV towards higher energies, as expected from theoretical calculations performed on stanene-graphene heterostructure. More importantly, similarities were observed in the optical response of the tin nanosheets and the semi-metallic allotropic form of tin, suggesting the possibility of stabilizing $\alpha$-Sn on Gr-Al$_2$O$_3$(0001) at room temperature. By comparing the optical properties of graphene and tin nanosheets, with and without a graphene buffer layer, our findings provide a promising route to modulate light absorption by active two-dimensional layers in the VIS and UV photon range, setting the scene for ultra-scaled photonic devices.

**CRediT author statement**

**Eleonora Bonaventura:** Investigation, Methodology, Formal analysis, Data Curation, Writing - Original Draft, Visualization. **Christian Martella:** Conceptualization, Methodology, Writing - Review & Editing, Supervision. **Salvatore Macis:** Formal analysis, Writing - Review & Editing. **Daya S. Dhungana:** Investigation, Writing - Review & Editing. **Simonas Krotkus:** Investigation. **Michael Heuken:** Investigation. **Stefano Lupi:** Writing - Review & Editing, Funding acquisition. **Alessandro Molle**: Conceptualization, Writing -Review & Editing, Supervision, Project administration, Funding acquisition. **Carlo Grazianetti**: Conceptualization, Methodology, Investigation, Writing - Review & Editing, Supervision.

**Acknowledgments**
Mario Alia (CNR-IMM) for technical assistance and Sara Ghomi (CNR-IMM) for the AFM analysis are acknowledged.

**Funding**
This research was funded by European Commission within the H2020 research and innovation programme under the ERC-COG 2017 grant no. 772261 "XFab" and partially supported by the Italian Ministry of University and Research (MUR) under the PRIN 2020 grant n. 2020RPEPNH "PHOTO".

## Appendix A - Supplementary Information

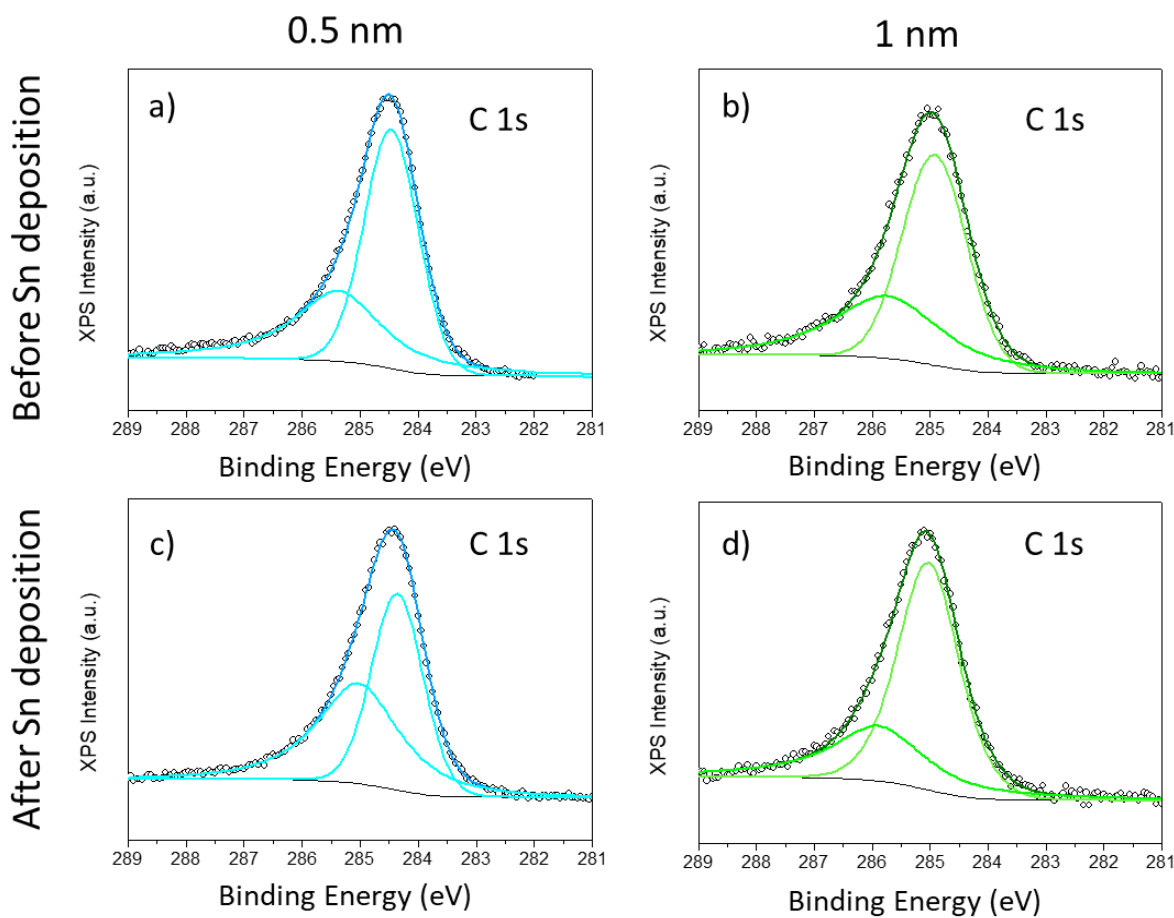

**Figure A.1:** XPS spectra of the C 1s core level measured before and after the deposition of 0.5 and 1 nm of tin. a) C 1s graphene ($sp^2$ carbon) and C-H core levels are placed at BE=284.46 and BE=285.38 eV, respectively, before Sn deposition. b) C 1s graphene ($sp^2$ carbon) and C-H core levels are placed at BE=284.92 and BE=285.74 eV, respectively, before Sn deposition. Both BEs are in good agreement with Ref. [19]. c) C 1s graphene ($sp^2$ carbon) and C-H core levels are placed at BE=284.36 and BE=285.04 eV, respectively, after 0.5 nm-thick Sn deposition. d) C 1s graphene ($sp^2$ carbon) and C-H core levels are placed at BE=285.03 and BE=285.89 eV, respectively, after 1 nm-thick Sn deposition.

| Case | 2D FWHM ($cm^{-1}$) | Error ($cm^{-1}$) | 2D/G | Error | D/G | Error |
| --- | --- | --- | --- | --- | --- | --- |
| Before cut | 38 | 3 | 2.7 | 0.6 | 0.22 | 0.05 |
| After cut - A | 40 | 4 | 2.5 | 0.4 | 0.21 | 0.03 |
| After cut - B | 39 | 4 | 2.5 | 0.4 | 0.24 | 0.07 |

**Table A.1:** Average value and standard deviation for 2D peak FWHM, 2D/G and D/G obtained in the three different areas indicated in Figure S2.

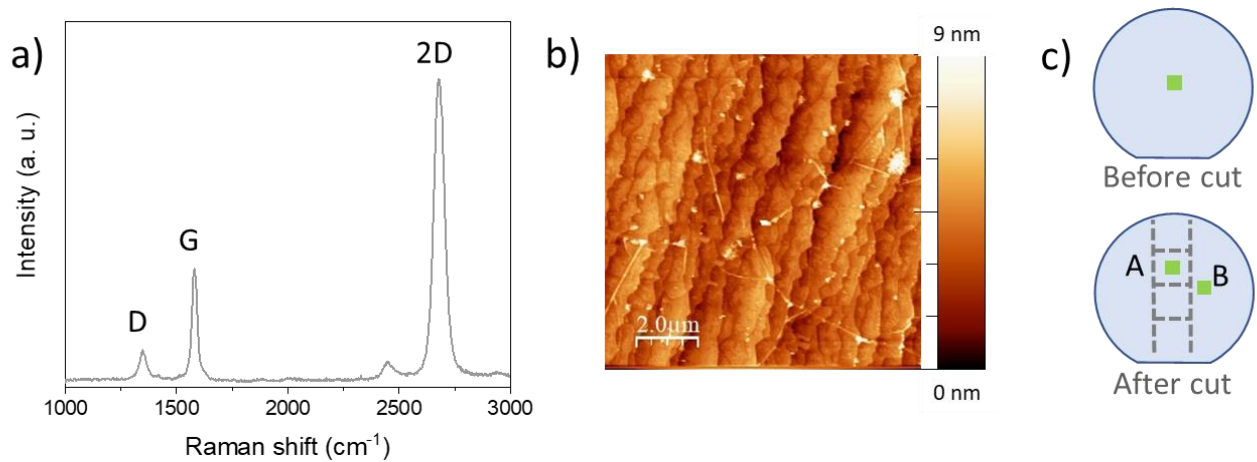

**Figure A.2:** a) Representative Raman spectrum of graphene on $Al_2O_3$ (0001). b) AFM line profile of terraces and wrinkles for graphene on $Al_2O_3$(0001) after the cutting process. c) Wafer sketches (non-scaled dimensions). The dashed grey lines (bottom) delimit the area of the slabs resulting from the cut. Green squares highlight the areas where Raman maps were acquired to determine initial and final conditions of graphene.

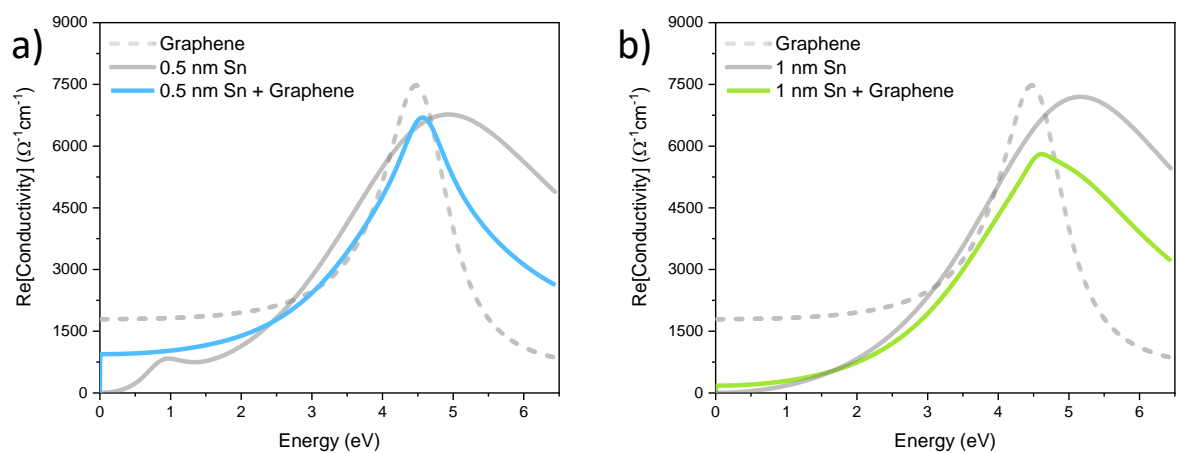

**Figure A.3:** a-b) Real part of the optical conductivity extracted from a bilayer model. In this case, in addition to the $Al_2O_3$ substrate, a single layer with a thickness equal to the sum of the thicknesses of the graphene and tin layers was considered. The optical response of the composite layer is compared with the optical response of the individual components discussed in the main text.

## References

[1] Y. Liu, N.O. Weiss, X. Duan, H.-C. Cheng, Y. Huang, X. Duan, Nat. Rev. Mater. 1 (2016) 16042.

[2] C. Grazianetti, A. Molle, in: Xenes, Elsevier, 2022, pp. 377–403.

[3] Y. Deng, Z. Luo, N.J. Conrad, H. Liu, Y. Gong, S. Najmaei, P.M. Ajayan, J. Lou, X. Xu, P.D. Ye, ACS Nano 8 (2014) 8292–8299.

[4] X. Chen, Y. Wu, Z. Wu, Y. Han, S. Xu, L. Wang, W. Ye, T. Han, Y. He, Y. Cai, N. Wang, Nat. Commun. 6 (2015) 7315.

[5] D.S. Dhungana, C. Grazianetti, C. Martella, S. Achilli, G. Fratesi, A. Molle, Adv. Funct. Mater. 31 (2021) 2102797.

[6] E. Bonaventura, D.S. Dhungana, C. Martella, C. Grazianetti, S. Macis, S. Lupi, E. Bonera, A. Molle, Nanoscale Horizons 7 (2022) 924–930.

[7] A. Castellanos-Gomez, M. Buscema, R. Molenaar, V. Singh, L. Janssen, H.S.J. van der Zant, G.A. Steele, 2D Mater. 1 (2014) 011002.

[8] R.A. Johnson, P.R. de la Houssaye, C.E. Chang, Pin-Fan Chen, M.E. Wood, G.A. Garcia, I. Lagnado, P.M. Asbeck, IEEE Trans. Electron Devices 45 (1998) 1047–1054.

[9] T. Baehr-Jones, A. Spott, R. Ilic, A. Spott, B. Penkov, W. Asher, M. Hochberg, Opt. Express 18 (2010) 12127.

[10] Z. Chen, Z. Liu, T. Wei, S. Yang, Z. Dou, Y. Wang, H. Ci, H. Chang, Y. Qi, J. Yan, J. Wang, Y. Zhang, P. Gao, J. Li, Z. Liu, Adv. Mater. 31 (2019) 1807345.

[11] C. Grazianetti, S. De Rosa, C. Martella, P. Targa, D. Codegoni, P. Gori, O. Pulci, A. Molle, S. Lupi, Nano Lett. 18 (2018) 7124–7132.

[12] C. Grazianetti, E. Bonaventura, C. Martella, A. Molle, S. Lupi, ACS Appl. Nano Mater. 4 (2021) 2351–2356.

[13] L. Matthes, O. Pulci, F. Bechstedt, New J. Phys. 16 (2014) 105007.

[14] R. John, B. Merlin, J. Phys. Chem. Solids 110 (2017) 307–315.

[15] J. Ahn, J.W. Rabalais, Surf. Sci. 388 (1997) 121–131.

[16] C. Niu, K. Shepherd, D. Martini, J. Tong, J.A. Kelber, D.R. Jennison, A. Bogicevic, Surf. Sci. 465 (2000) 163–176.

[17] P. V. Pham, S.C. Bodepudi, K. Shehzad, Y. Liu, Y. Xu, B. Yu, X. Duan, Chem. Rev. 122 (2022) 6514–6613.

[18] N. Mishra, S. Forti, F. Fabbri, L. Martini, C. McAleese, B.R. Conran, P.R. Whelan, A. Shivayogimath, B.S. Jessen, L. Buß, J. Falta, I. Aliaj, S. Roddaro, J.I. Flege, P. Bøggild, K.B.K. Teo, C. Coletti, Small 15 (2019) 1904906.

[19] Z. Chen, C. Xie, W. Wang, J. Zhao, B. Liu, J. Shan, X. Wang, M. Hong, L. Lin, L. Huang, X. Lin, S. Yang, X. Gao, Y. Zhang, P. Gao, K.S. Novoselov, J. Sun, Z. Liu, Sci. Adv. 7 (2021) 115.

[20] J. Li, M. Chen, A. Samad, H. Dong, A. Ray, J. Zhang, X. Jiang, U. Schwingenschlögl, J. Domke, C. Chen, Y. Han, T. Fritz, R.S. Ruoff, B. Tian, X. Zhang, Nat. Mater. 21 (2022) 740–747.

[21] A.T. Hoang, A.K. Katiyar, H. Shin, N. Mishra, S. Forti, C. Coletti, J.-H. Ahn, ACS Appl. Mater. Interfaces 12 (2020) 44335–44344.

[22] G. Piccinini, S. Forti, L. Martini, S. Pezzini, V. Miseikis, U. Starke, F. Fabbri, C. Coletti, 2D Mater. 7 (2020) 014002.

[23] J. Sone, T. Yamagami, K. Nakatsuji, H. Hirayama, Jpn. J. Appl. Phys. 55 (2016) 035502.

[24] G. Li, L. Zhang, W. Xu, J. Pan, S. Song, Y. Zhang, H. Zhou, Y. Wang, L. Bao, Y. Zhang, S. Du, M. Ouyang, S.T. Pantelides, H. Gao, Adv. Mater. 30 (2018) 1804650.

[25] Z. Ben Jabra, M. Abel, F. Fabbri, J.-N. Aqua, M. Koudia, A. Michon, P. Castrucci, A. Ronda, H. Vach, M. De Crescenzi, I. Berbezier, ACS Nano 16 (2022) 5920–5931.

[26] L. Wu, P. Lu, J. Bi, C. Yang, Y. Song, P. Guan, S. Wang, Nanoscale Res. Lett. 11 (2016).

[27] X. Chen, R. Meng, J. Jiang, Q. Liang, Q. Yang, C. Tan, X. Sun, S. Zhang, T. Ren, Phys. Chem. Chem. Phys. 18 (2016) 16302–16309.

[28] H. Wu, J. Tang, Q. Liang, B. Shi, Y. Niu, J. Si, Q. Liao, W. Dou, Appl. Phys. Lett. 115 (2019) 141601.

[29] A. Molle, G. Faraone, A. Lamperti, D. Chiappe, E. Cinquanta, C. Martella, E. Bonera, E. Scalise, C. Grazianetti, Faraday Discuss. 227 (2021) 171–183.

[30] H. Song, J. Yao, Y. Ding, Y. Gu, Y. Deng, M.-H. Lu, H. Lu, Y.-F. Chen, Adv. Eng. Mater. 21 (2019) 1900410.

[31] M. Li, L. Zheng, M. Zhang, Y. Lin, L. Li, Y. Lu, G. Chang, P.J. Klar, Y. He, Appl. Surf. Sci. 466 (2019) 765–771.

[32] X. Zheng, J.-F. Zhang, B. Tong, R.-R. Du, 2D Mater. 7 (2020) 011001.

[33] A.C. Ferrari, D.M. Basko, Nat. Nanotechnol. 8 (2013) 235–246.

[34] J.E. Lee, G. Ahn, J. Shim, Y.S. Lee, S. Ryu, Nat. Commun. 3 (2012) 1024.

[35] H.J. Song, M. Son, C. Park, H. Lim, M.P. Levendorf, A.W. Tsen, J. Park, H.C. Choi, Nanoscale 4 (2012) 3050.

[36] H. Wördenweber, S. Karthäuser, A. Grundmann, Z. Wang, S. Aussen, H. Kalisch, A. Vescan, M. Heuken, R. Waser, S. Hoffmann-Eifert, Sci. Rep. 12 (2022) 18743.

[37] A. Das, S. Pisana, B. Chakraborty, S. Piscanec, S.K. Saha, U. V Waghmare, K.S. Novoselov, H.R. Krishnamurthy, A.K. Geim, A.C. Ferrari, A.K. Sood, Nat. Nanotechnol. 3 (2008) 210–215.

[38] D. Yoon, Y.-W. Son, H. Cheong, Phys. Rev. Lett. 106 (2011) 155502.

[39] F. Fabbri, M. Scarselli, N. Shetty, S. Kubatkin, S. Lara-Avila, M. Abel, I. Berbezier, H. Vach, M. Salvato, M. De Crescenzi, P. Castrucci, Surfaces and Interfaces 33 (2022) 102262.

[40] T.M.G. Mohiuddin, A. Lombardo, R.R. Nair, A. Bonetti, G. Savini, R. Jalil, N. Bonini, D.M. Basko, C. Galiotis, N. Marzari, K.S. Novoselov, A.K. Geim, A.C. Ferrari, Phys. Rev. B 79 (2009) 205433.

[41] A.B. Kuzmenko, Rev. Sci. Instrum. 76 (2005) 083108.

[42] A.B. Kuzmenko, E. van Heumen, F. Carbone, D. van der Marel, Phys. Rev. Lett. 100 (2008) 117401.

[43] A. Perucchi, L. Baldassarre, C. Marini, P. Postorino, F. Bernardini, S. Massidda, S. Lupi, Phys. Rev. B 86 (2012) 035114.

[44] K.F. Mak, M.Y. Sfeir, Y. Wu, C.H. Lui, J.A. Misewich, T.F. Heinz, Phys. Rev. Lett. 101 (2008) 196405.

[45] K.F. Mak, J. Shan, T.F. Heinz, Phys. Rev. Lett. 106 (2011) 046401.

[46] R.A. Carrasco, C.M. Zamarripa, S. Zollner, J. Menéndez, S.A. Chastang, J. Duan, G.J. Grzybowski, B.B. Claflin, A.M. Kiefer, Appl. Phys. Lett. 113 (2018) 232104.

[47] K. Takeuchi, S. Adachi, J. Appl. Phys. 105 (2009) 073520.

[48] X. Dong, L. Zhang, M. Yoon, P. Zhang, 2D Mater. 8 (2021) 045003.